\documentclass[conference]{IEEEtran}
\IEEEoverridecommandlockouts
\usepackage{cite}
\usepackage{amsmath,amssymb,amsfonts}
\usepackage{algorithmic}
\usepackage{graphicx}
\usepackage{multirow}   
\usepackage{textcomp}
\usepackage{xcolor}
\usepackage{subcaption}
\def\BibTeX{{\rm B\kern-.05em{\sc i\kern-.025em b}\kern-.08em
    T\kern-.1667em\lower.7ex\hbox{E}\kern-.125emX}}
\makeatletter
\newcommand{\linebreakand}{%
  \end{@IEEEauthorhalign}
  \hfill\mbox{}\par
  \mbox{}\hfill\begin{@IEEEauthorhalign}
}
\makeatother
\begin{document}

\title{WiRD-Gest: Gesture Recognition In The Real World Using Range-Doppler Wi-Fi Sensing on COTS Hardware\\

\thanks{Part of this work received funding from the EC Horizon Europe SNS JU projects 6G-SENSES (GA 101139282) and MultiX (GA 101192521).}
}

\author{
    \IEEEauthorblockN{Jessica Sanson}
    \IEEEauthorblockA{\textit{Intel Deutschland GmbH} \\
    Munich, Germany \\
    jessica.sanson@intel.com}
    \and
    \IEEEauthorblockN{Rahul C. Shah}
    \IEEEauthorblockA{\textit{Intel Labs} \\
    Santa Clara, CA, USA \\
    rahul.c.shah@intel.com}
    \and
    \IEEEauthorblockN{Yazhou Zhu}
    \IEEEauthorblockA{\textit{Intel Deutschland GmbH} \\
    Munich, Germany \\
    yazhou.zhu@intel.com}
    \linebreakand 
    \IEEEauthorblockN{Rafael Rosales}
    \IEEEauthorblockA{\textit{Intel Deutschland GmbH} \\
    Munich, Germany \\
    }
    \and
    \IEEEauthorblockN{Valerio Frascolla}
    \IEEEauthorblockA{\textit{Intel Deutschland GmbH} \\
    Munich, Germany \\
    valerio.frascolla@intel.com}
}
\maketitle

\begin{abstract}
Wi-Fi sensing has emerged as a promising technique for gesture recognition, yet its practical deployment is hindered by environmental sensitivity and device placement challenges. To overcome these limitations we propose Wi-Fi Range and Doppler (WiRD)-Gest, a novel system that performs gesture recognition using a single, unmodified Wi-Fi transceiver on a commercial off-the-shelf (COTS) laptop. The system leverages an monostatic full duplex sensing pipeline capable of extracting Range-Doppler (RD) information. Utilizing this, we present the first benchmark of deep learning models for gesture recognition based on monostatic sensing. The key innovation lies in how monostatic sensing and spatial (range) information fundamentally transforms accuracy, robustness and generalization compared to prior approaches. We demonstrate excellent performance in crowded, unseen public spaces with dynamic interference and additional moving targets even when trained on data from controlled environments only. These are scenarios where prior Wi-Fi sensing approaches often fail, however, our system suffers minor degradation. The WiRD-Gest benchmark and dataset will also be released as open source.
\end{abstract}


\section{Introduction}
\label{sec:intro}
Recent advances in Wi-Fi sensing \cite{b0, Wi-Fi-Sensing} have demonstrated capabilities in gesture detection by analyzing Channel State Information (CSI) patterns. By leveraging perturbations in  CSI, researchers have developed systems capable of recognizing a wide range of human activities and gestures~\cite{b9, b7,SenseFi,WiDir,WiBorder,Multisense}. However, most state-of-the-art methods are based on bistatic sensing, where models are trained on CSI amplitude and phase variations obtained along a communication link between two Wi-Fi endpoints~\cite{SenseFi}. A recent survey on Machine Learning (ML) techniques for Wi-Fi CSI-based Recognition and Sensing can be found in \cite{Survey}.

While effective in controlled, single-user environments, these approaches are fundamentally limited by their inability to extract precise spatial information~\cite{UWB-Fi}, making them highly susceptible to environmental changes, device deployments and movements from other targets~\cite{Widar3,WiGRUNT}. Without range information, a system cannot spatially distinguish a user's intended gesture from background motion, leading to a drastic drop in performance.

While sophisticated techniques have been developed to improve cross-domain robustness—such as the attention mechanisms in WiGRUNT~\cite{WiGRUNT} or the domain-independent Body-Coordinate Velocity Profile (BVP) in Widar3.0~\cite{Widar3}—they still do not resolve the fundamental spatial ambiguity between a gesture in the foreground and background movements. Even the comprehensive SenseFi benchmark~\cite{SenseFi}, which evaluates a variety of deep learning models, or the ambitious and promising approach of the foundation model Wisensenet\cite{WiSenseNet}, focus on CSI data that lacks this crucial spatial dimension. To combat this, \cite{UWB-Fi} uses spatial information as input data; however, changes to the hardware are needed to deliver the anticipated improvements. Consequently, there is a critical gap in the literature to achieve reliable gesture recognition on commercial
off-the-shelf (COTS) devices in uncontrolled, multi-person environments.

Recent works have begun to explore monostatic sensing over Wi-Fi to overcome these limitations. For instance, \cite{mmwave-monostatic} proposes a first validated implementation of an
Orthogonal Frequency Division Multiplexing (OFDM)-based mmWave Wi-Fi integrated Sensing and Communication (ISAC) system for monostatic
human sensing at 60 GHz, whereas
{\itshape ISAC-Fi} \cite{chen2024isacfi} demonstrates a prototype Wi-Fi device with self-interference cancellation enabling monostatic sensing under standard communication workloads. Similarly, a recent Software Defined Radio (SDR)-based implementation \cite{Kristensen2025monostatic} shows stable, long-duration human motion sensing (e.g.,\ breathing) up to 10 m under non-line-of-sight conditions. However, these solutions focus on mmWave scenarios, or rely on custom prototypes and SDR rather than standard commercial devices. 
\begin{figure}[htbp]
    \centering
    \begin{subfigure}[b]{0.51\linewidth}
        \centering
        \includegraphics[width=\linewidth]{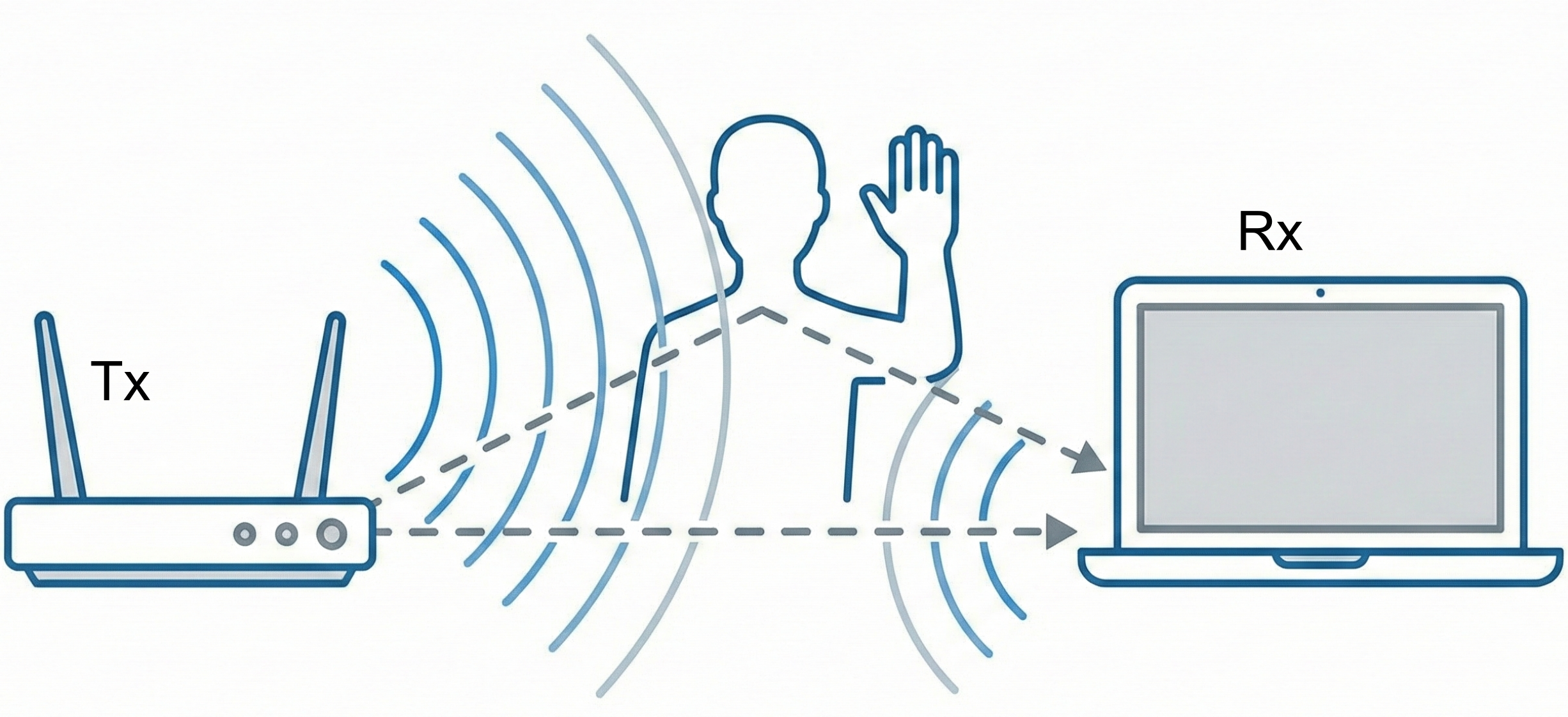}
        \caption{Bistatic  Wi-Fi gesture recognition topology (traditional).}
        \label{fig:bistatic_gesture}
    \end{subfigure}
    \hfill
    \begin{subfigure}[b]{0.47\linewidth}
        \centering
        \includegraphics[width=\linewidth]{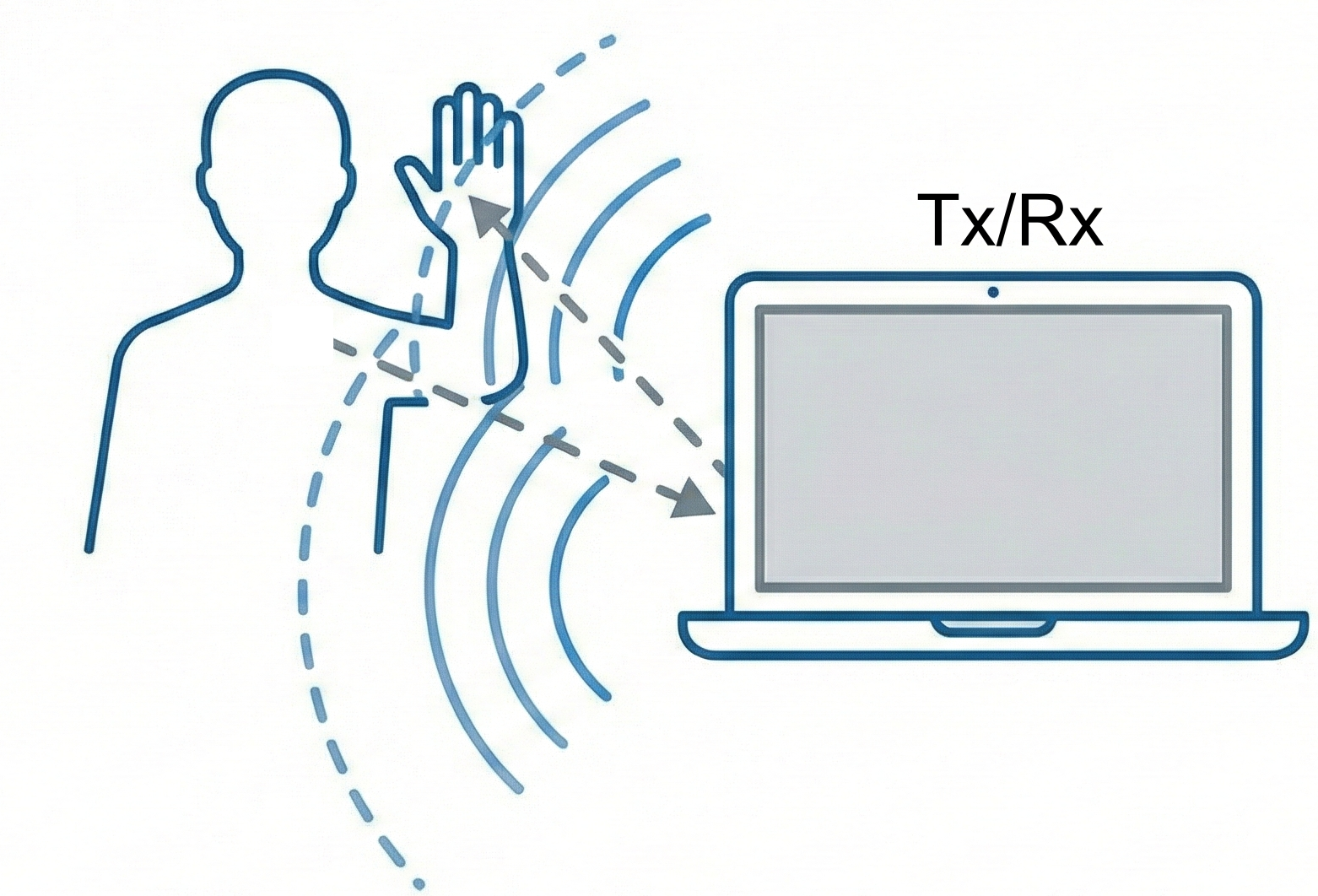}
        \caption{Monostatic Wi-Fi gesture recognition topology (proposed).}
        \label{fig:monostatic_gesture}
    \end{subfigure}
    \caption{Visual comparison of Wi-Fi sensing scenarios. (a) Bistatic sensing requires a separate transmitter and receiver. (b) The proposed system captures gestures using a single device.}
    \label{fig:sensing_scenarios}
\end{figure}

\begin{figure}[htbp]
    \centering
    \begin{subfigure}[b]{0.9\linewidth}
        \centering
        \includegraphics[width=\linewidth]{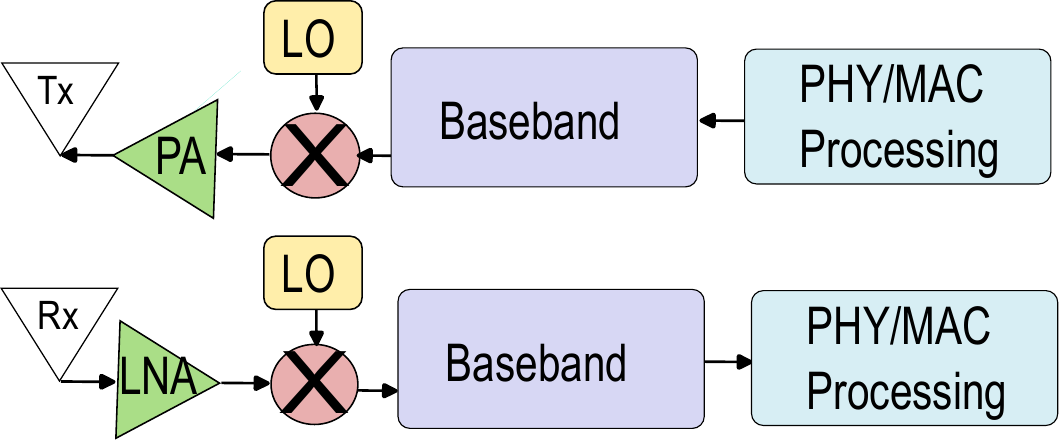} %
        \caption{Bistatic transceiver architecture.}
        \label{fig:bistatic_block}
    \end{subfigure}
    
    \begin{subfigure}[b]{0.9\linewidth}
        \centering
        \includegraphics[width=\linewidth]{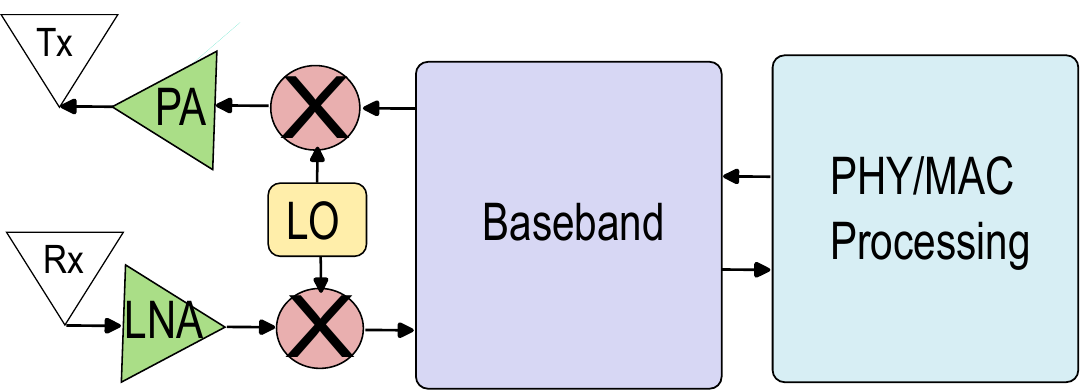} %
        \caption{Monostatic transceiver architecture.}
        \label{fig:monostatic_block}
    \end{subfigure}
    \caption{Comparison of transceiver architectures. (a) Bistatic mode separates Tx and Rx on different hardware. (b) Monostatic mode  utilizes a shared LO and Baseband processor on a single device.}
    \label{fig:architectures}
\end{figure}

To address this gap, our gesture recognition system builds upon the full-duplex monostatic pipeline we introduced in~\cite{sansonRange}. While we previously demonstrated that this architecture can effectively detect human presence in~\cite{presence_percom}, this work extends the capability to exploit Range-Doppler maps for fine-grained gesture recognition. Partially funded by the MultiX~\cite{MultiX} and 6G-SENSES~\cite{6G-SENSES} projects, our approach uses CSI data captured directly on a single device. Figure~\ref{fig:sensing_scenarios} illustrates the physical distinction: unlike traditional bistatic systems that rely on separated Tx/Rx pairs (Fig.~\ref{fig:bistatic_gesture}), our approach captures gestures monostatically on a single laptop (Fig.~\ref{fig:monostatic_gesture}), utilizing the transceiver architecture detailed in Figure~\ref{fig:architectures}.

While bistatic sensing requires synchronization between distinct hardware units, the proposed monostatic architecture (Fig. \ref{fig:monostatic_block}) shares the Local Oscillator (LO) and Baseband processing between the Tx and Rx chains, enabling self-contained sensing.

Crucially, our system maintains simultaneous normal Wi-Fi communication and Internet connectivity with no Network Interface Card (NIC) modification. This method allows the extraction of high-fidelity Range-Doppler (RD) information from a COTS laptop.  Using these RD maps as input data for the models, we present a comprehensive ML benchmark specifically designed for Wi-Fi monostatic sensing data. Our primary contribution is to demonstrate that the inclusion of range information creates a paradigm shift in sensing robustness. Most critically, we validate this by deploying our system in a crowded public environment without any site-specific data collection or adaptation — a scenario not addressed by existing literature.

Our key contributions are:

\begin{itemize}
\item The first benchmark of deep learning architectures on Wi-Fi monostatic sensing based on RD data on commercial devices.
\item The first demonstration of Wi-Fi gesture recognition with high accuracy in uncontrolled public environments with significant background motion.
\item The first open source public dataset based on monostatic Wi-Fi sensing on commercial devices ~\cite{sanson2026wirdgest_dataset}.
\end{itemize}

\section{System Model}
\label{sec:background}

\subsection{Range-Doppler Map Generation}
The proposed monostatic approach utilizes a COTS Wi-Fi NIC with dedicated Tx and Rx antennas on a single device. To keep this paper self-contained, we briefly recapitulate the signal model already introduced in~\cite{sansonRange, presence_percom}. We define the frequency-domain channel matrix of the Long Training Field (LTF)—comprising $N$ subcarriers and $M$ OFDM frames reflected by $K$ targets—at subcarrier $n$ and frame $m$ as \cite{sanson2020ofdm}:
\begin{equation}
D(m,n) = \sum_{k=1}^{K} e^{j2\pi T m f_{D,k}} e^{-j2\pi n \Delta f \tau_k} + \tilde{\eta}
\end{equation}
where $f_{D,k}$ is the Doppler shift, $\tau_k$ is the delay for target $k$, $T$ is the frame interval, $\Delta f$ is the subcarrier spacing, and $\tilde{\eta}$ is additive white Gaussian noise (AWGN). RD maps are obtained via a two-dimensional DFT (2D-FFT) across frames and subcarriers.
\subsection{Synchronization}
Real commercial Wi‑Fi hardware (COTS NICs) is not designed for full‑duplex radar‑like operation: hardware asynchronization (e.g., unsynchronized time/phase clocks) and strong Tx–Rx coupling (self‑interference) pose serious challenges. To address these, we apply the algorithm described in~\cite{sansonRange}, which performs delay and phase alignment and  self-interference mitigation. Our implementation  performs the follows steps:

\textbf{Delay calibration:} We first perform coarse synchronization by computing the cross-correlation between the known training-symbol sequence, yielding $l_{\text{coarse}} = \arg\max_{l} C(l)$. Then, a refined correlation (upsampled by factor $U$) gives the fine delay $l_{\text{fine}} = \arg\max_{l} C_{\text{fine}}(l)$ \cite{sansonRange}. The total delay correction is $l_{\text{eff}} = l_{\text{coarse}} + l_{\text{fine}}$

\textbf{Phase synchronization:} For each frame $m$, the average phase is computed as
$
\theta_m = \angle \left(\frac{1}{N} \sum_{n=1}^{N} D(m,n)\right)
$. The reference phase $\phi_{m-1}$ is the average over the previous $H$ frames. The phase difference is $\Delta\theta_m = \phi_{m-1} - \theta_m$, quantized in steps of $\delta$ \cite{sansonRange}, with
\begin{equation}
\text{fix}_m = \operatorname{round}\left(\frac{\Delta \theta_m}{\delta}\right) \cdot \delta
\end{equation}
The phase of all subcarriers is then corrected by 
\begin{equation}
D(m,n) \leftarrow D(m,n) \cdot \exp\left(j\,\text{fix}_m\right)
\end{equation}

\begin{figure*}[t]
    \centering
    \begin{subfigure}[b]{0.47\linewidth}
        \includegraphics[width=\linewidth]{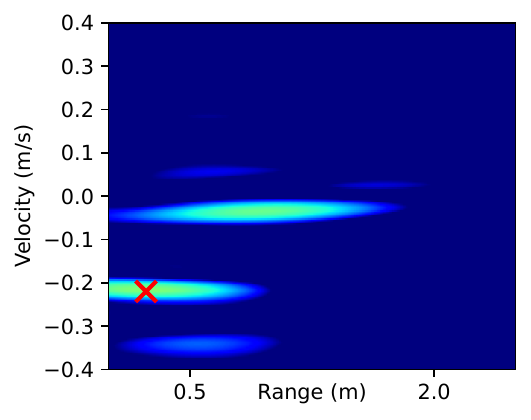}
        \caption{Up-Down gesture}
        \label{fig:rd_updown}
    \end{subfigure}
    \hfill
    \begin{subfigure}[b]{0.465\linewidth}
        \includegraphics[width=\linewidth]{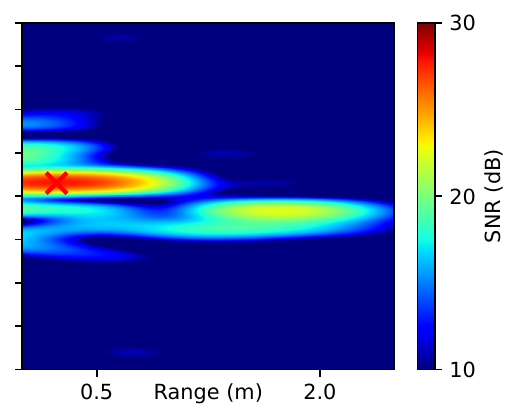}
        \caption{Rotate gesture}
        \label{fig:rd_rotate}
    \end{subfigure}
    \vspace{-0.8em}
    \caption{
    RD maps (single frame) from the public space location showing (a) Up-Down and (b) Rotate gestures. Note the secondary moving targets in the background in both cases.
    }
    \label{fig:range_doppler}
\end{figure*}

\textbf{Self-interference cancellation:} The Tx/Rx coupling is removed by subtracting the mean across all $M$ frames $\hat{D}(m,n) = D(m,n) - \frac{1}{M} \sum_{m=0}^{M-1} D(m,n)$ \cite{sansonRange}. This synchronization and interference suppression enables high-fidelity RD map estimation, as shown in Fig.~\ref{fig:range_doppler}. 

\section{Models and Training}
To provide a comprehensive evaluation of model performance and generalization on RD data, we benchmark a diverse set of state-of-the-art and classical deep learning architectures:

\textbf{CNN2D + RNN/GRU/LSTM:} These models process each RD frame independently with 2D convolutions for spatial feature extraction, then use a recurrent module for temporal sequence modeling. We implement variants using vanilla Recurrent Neural Network (RNN)~\cite{RNN}, Gated Recurrent Unit (GRU)~\cite{GRU}, and Long Short-Term Memory (LSTM)~\cite{LSTM}. GRU and LSTM offer improved capacity for modeling long-term dependencies versus vanilla RNN~\cite{GRU,LSTM}, with LSTM generally requires more computation. These architectures are lightweight and well-suited to real-time applications.

\noindent\textbf{3D CNN:} 3D Convolutional Neural Networks (CNNs) perform spatiotemporal convolutions directly over stacked RD frames, learning joint motion patterns across space and time~\cite{C3D,Molchanov2016}. We evaluate standard 3D CNNs as well as deeper residual variants (ResNet3D-18/34/50)~\cite{ResNet}. The residual architecture improves trainability for deeper networks~\cite{ResNet} yielding high accuracy on video gesture recognition. 

\noindent\textbf{Video Transformers:} We adapt recent video transformer models—ViViT~\cite{ViViT} and TimeSformer~\cite{TimeSformer}—for RD gesture sequences. Both operate on spatiotemporal patches: ViViT factorizes spatial and temporal attention for efficiency~\cite{ViViT}, while TimeSformer divides space-time attention within each block~\cite{TimeSformer}. These architectures enable modeling of complex, long-range dependencies and are competitive with 3D CNNs, with a distinct trade-off between accuracy and inference cost.

\subsection{Input Representation and Data Pipeline}
We model each gesture as a short RD clip/video. Each RD frame is converted from raw Signal-to-Noise Ratio (SNR) values to a normalized tensor by clipping to $[5, 40]$ dB and scaling to $[0, 1]$. Frames are resized to a square target resolution (e.g., $64 \times 64$) with a single channel. The range and Doppler axes are interpolated to achieve cell sizes of $0.93$ cm and $0.015$ m/s, respectively, generating RD spectrograms covering a range of $0$ to $0.63$ m and velocities of $\pm 0.45$ m/s. Clips are segmented into fixed-length sequences using a sliding window. Each sequence is aligned with annotated gesture labels, and only complete gesture windows are included. Each clip consists of $32$ RD frames, for a total window duration of $3.2$ s, and contains  one complete gesture plus a short pre/post buffer. Tensors are fed to models that either (i) process frames independently then aggregate temporally (2D-CNN $\rightarrow$ RNN/GRU/LSTM), or (ii) learn joint spatiotemporal features directly (3D CNN / 3D-ResNet / Video Transformer).

\subsection{Dataset and Data Collection}
\textbf{Hardware Setup:} All data were collected using a single commercial laptop (Lenovo ThinkPad L14) equipped with an Intel Wi-Fi 6E AX211 NIC, operating in channel 79 (6.345 GHz center frequency), with a bandwidth of $160$ MHz, $512$ subcarriers, and a frame rate of $40$ Hz ($\Rightarrow \pm0.4$ m/s unambiguous velocity). The system provides range and Doppler resolutions of $0.93$ cm (160 MHz bandwidth) and $0.03$ m/s, respectively.


\textbf{Dataset:} We collected a comprehensive dataset of $191000$ frame samples ($77500$ gesture frames) to train and evaluate our models in controlled settings. Five distinct users participated in data collection, each instructed to perform five common gestures: (1) Push \& Pull - hand toward and away from PC, (2) Slide - side to side motion, (3) Up-Down - up and down motion, (4) Double Pulse - two open hands in front of PC, and (5) Double Rotate - rotate twice clockwise in front of PC. To ensure some environmental diversity, data were collected in two different indoor locations: data from two users were collected in a furnished office room (location A), while the other three users' data were collected in an open meeting room (location B). For each gesture, every user performed $20$ repetitions for the training set, plus $8$ repetitions for the test set. This resulted in a total dataset of $720$ gesture instances where 30\% was used for validation and testing (split randomly into 15\% validation and 15\% testing).

\noindent\textbf{Unseen public-space validation.} To assess robustness in the real world, one user also performed the gestures in a third location (location C) - a public space (cafe) characterized by dense foot traffic and monostatic Wi-Fi devices. This data is never used for training or model selection and serves exclusively as an out-of-domain test set.

\section{Experiment}
For all experiments, we report accuracy, macro-F1 score, number of model parameters, and GFLOPS to characterize both recognition performance and computational complexity. Models are evaluated in three settings: \emph{in-domain} (random per-instance split, locations A and B, all users), \emph{cross-user} (leave-one-user-out), and \emph{unseen public space/cross-location} (training on locations A and B, testing on public location C). 

\noindent\textbf{Training Configuration:} The designs of our models are based on established practices in recent literature on Wi-Fi and video-based gesture recognition, as well as on our own empirical studies. For each  model we conduct systematic evaluations of architectural configurations to identify the variants achieving optimal performance. In addition, we explore a wide range of hyperparameters to ensure fair comparison and robust generalization. All models are trained with cross-entropy loss, Adam optimizer, learning rate and weight decay of \(1\times10^{-3}\), dropout of 0.2, batch size of \(16\), and up to \(100\) epochs. To ensure statistical reliability, each model was trained five independent times, all reported metrics reflect the best-performing run.

\subsection{Overall Model Comparison}
\begin{figure}[t]
    \centering
    \includegraphics[width=0.95\linewidth]{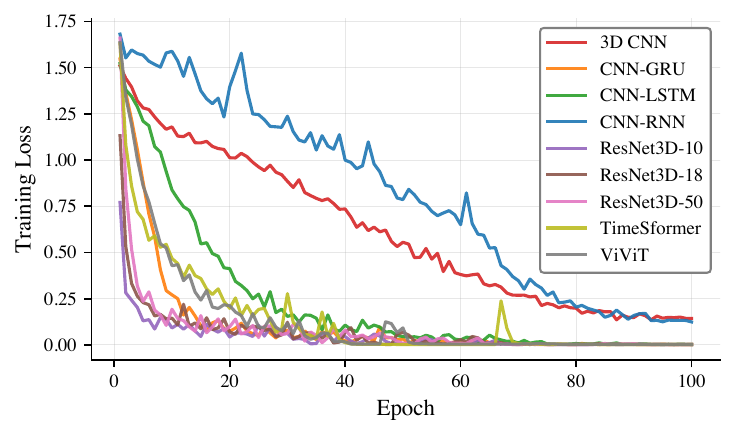}
    \caption{Training loss curves for different models}
   \label{fig:comparative_training_loss}
\end{figure}

\begin{table}[t]
\centering
\caption{Performance comparison -- Wi-Fi gesture recognition}
\label{tab:main}
\footnotesize
\setlength{\tabcolsep}{3.5pt}
\begin{tabular}{lcccc}
\hline
\textbf{Model} & \textbf{Accuracy (\%)} & \textbf{F1-Macro (\%)} & \textbf{Parameters} & \textbf{GFLOPS} \\
\hline
3D CNN             & 78.80 & 78.99 & 344,581     & 3.77 \\
CNN-GRU            & \textbf{95.18} & \textbf{95.18} & 537,669     & 1.89 \\
CNN-LSTM           & 91.81 & 91.81 & 587,077     & 1.89 \\
CNN-RNN            & 77.83 & 78.12 & 438,853     & 1.88 \\
ResNet3D-10        & 92.53 & 92.54 & 14,358,085  & 3.75 \\
ResNet3D-18        & 91.81 & 91.86 & 33,162,565  & 7.15 \\
ResNet3D-50        & 89.64 & 89.71 & 46,165,317  & 9.49 \\
TimeSformer        & 81.45 & 81.36 & 202,629     & 0.08 \\
ViViT              & 82.89 & 83.04 & \textbf{144,901}     & \textbf{0.08} \\
\hline
\end{tabular}
\end{table}

Fig.~\ref{fig:comparative_training_loss} shows the training loss curves for all models and
Table~\ref{tab:main} compares the results for the in-domain evaluation. The CNN-GRU architecture achieves the highest accuracy (95.2\%), with ResNet3D-10 as the next best performer. In general, models combining CNN-based spatial feature extraction with temporal modeling (GRU/LSTM) consistently outperform standalone 3D CNNs. Unlike 3D CNNs, which process time as a static block and can blur sequential movements, temporal models track the signal step-by-step. They maintain a memory of the ongoing changes in speed and direction. Because human gestures unfold sequentially, this continuous tracking is naturally better suited to capture the physical movement and reject random background noise. The relatively lower performance of CNN-RNN architectures aligns with prior reports of these models struggling with long sequence dependencies. Notably, the ResNet3D family demonstrates competitive results, being the first to converge on the training process, underlining the value of residual connections in video models applied to Wi-Fi RD data. Transformer-based models (TimeSformer, ViViT) deliver moderate performance, which we attribute to limited dataset size, howerer their efficiency and compactness outperform much larger 3D CNNs on a per-parameter basis.

As shown in Table~\ref{tab:pergesture}, per-class evaluation for the best-performing CNN-GRU model indicates high and consistent recognition accuracy across all gestures, with a small decrease observed for the side gesture. We hypothesize this is due to the hand’s profile orientation resulting in a smaller radar cross-section, increasing susceptibility to noise.

\begin{figure*}[t]
    \centering
    \begin{subfigure}[b]{0.45\linewidth}
        \includegraphics[width=\linewidth]{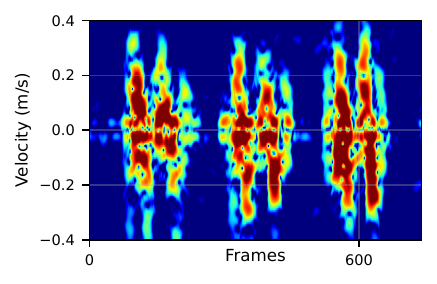}
        \caption{Controlled: With range}
    \end{subfigure}
    \hfill
    \begin{subfigure}[b]{0.45\linewidth}
        \includegraphics[width=\linewidth]{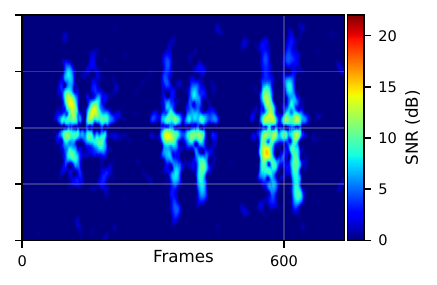}
        \caption{Controlled: No range}
    \end{subfigure}

    \begin{subfigure}[b]{0.45\linewidth}
        \includegraphics[width=\linewidth]{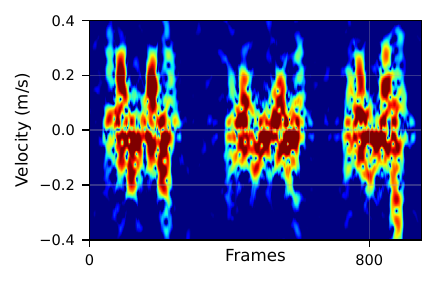}
        \caption{Public: With range}
    \end{subfigure}
    \hfill
    \begin{subfigure}[b]{0.45\linewidth}
        \includegraphics[width=\linewidth]{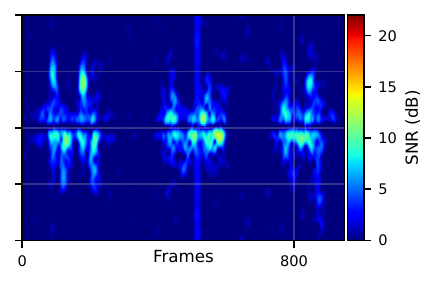}
        \caption{Public: No range}
    \end{subfigure}
    \vspace{-0.7em}
    \caption{
    Velocity spectrograms for three consecutive rotate gestures, comparing effect of using range to filter out background motion on Doppler information. Top: controlled environment and bottom: public space. Left: With range filtering ($> 1m$) and right: Without range filtering (all subcarriers). 
    }
    \label{fig:velocity_spectrograms}
\end{figure*}



\begin{table}[t]
\centering
\caption{Per-gesture accuracy and macro-F1 for the CNN-GRU model.}
\label{tab:pergesture}
\footnotesize
\begin{tabular}{lccccc}
\hline
             & \textbf{Push-Pull} & \textbf{Slide} & \textbf{Up-Down} & \textbf{Pulse} & \textbf{Rotate} \\
\hline
\textbf{Acc. (\%)} & 97.59 & 90.36 & 95.18 & 95.18 & 97.59 \\
\textbf{F1 (\%)}   & 98.78 & 91.46 & 94.05 & 94.61 & 97.01 \\
\hline
\end{tabular}
\end{table}

 The robustness of the models to user and environment variations is summarized in Table~\ref{tab:generalization}. In cross-user evaluation, the CNN-GRU architecture maintains high accuracy (89.4\%), demonstrating strong generalization to unseen users.

\begin{table}[t]
\centering
\caption{User and location (public space) independency accuracy and F1 for Wi-Fi gesture recognition models}
\label{tab:generalization}
\footnotesize
\setlength{\tabcolsep}{6pt}
\begin{tabular}{lcc|cc}
\hline
\multirow{2}{*}{\textbf{Model}} 
& \multicolumn{2}{c|}{\textbf{User Independence}} 
& \multicolumn{2}{c}{\textbf{Public Space}} \\
& \textbf{Acc. (\%)} & \textbf{F1 (\%)} 
& \textbf{Acc. (\%)} & \textbf{F1 (\%)} \\
\hline
3D CNN         & 77.18 & 77.91 & 60.81 & 63.12 \\
CNN-GRU        & \textbf{90.29} & \textbf{89.84} & \textbf{90.54} & \textbf{87.77} \\
CNN-LSTM       & 89.32 & 88.31 & 83.78 & 82.89 \\
CNN-RNN        & 68.45 & 66.82 & 79.73 & 76.79 \\
ResNet3D-10    & 86.41 & 90.35 & 79.73 & 77.63 \\
ResNet3D-18    & 84.47 & 89.03 & 82.43 & 82.27 \\
ResNet3D-50    & 81.55 & 85.33 & 79.73 & 80.76 \\
TimeSformer    & 81.55 & 81.62 & 75.68 & 75.67 \\
ViViT          & 73.30 & 72.65 & 79.73 & 81.01 \\
\hline
\end{tabular}
\end{table}

\noindent\textbf{Public-space validation.} Finally, to assess model robustness in realistic and interference-prone environments, we perform cross-location (public space) evaluation with the presence of multiple users and  wireless devices (laptops/mobiles).  During the experiments, at least three people were seated less than \hbox{1.5 m} from the experiment, and around ten people were within \hbox{5 m}. Additionally, multiple people frequently walked within \hbox{2 m} of the experimental area, Figure~\ref{fig:range_doppler} shows two examples of RD maps, where a gesture can be seen performed at 20 cm together with background moving targets at 1 and 2 m.

In Fig.~\ref{fig:velocity_spectrograms}, we compare the velocity spectrograms of consecutive rotate gestures under controlled and public space environments, with and without using range information to filter out background motion. Using range information to filter out distant gates ($>1$ $m$) significantly improves the SNR. In the public space, using Doppler processing only (based on bistatic sensing/mean over all subcarriers) results in visible degradation due to interference from bystanders. In contrast, range filtering robustly suppresses background noise and enables reliable gesture detection for practical Wi-Fi sensing applications in the real world.

Table~\ref{tab:generalization} presents  the models performance comparison in this real-world scenario. CNN-GRU model consistently achieves the highest accuracy (90.5\%) in the public space. The additional range dimension  appears to  support  generalization on temporal models even in complex and uncontrolled environments. To the best of our knowledge, this work is the first to validate gesture recognition models with COTS devices in a genuinely unconstrained public setting.

In contrast, models such as ResNet3D and 3D CNN experience performance degradation—exceeding 10\% loss in accuracy—under these conditions. This degradation can be attributed to the inherent limitations of fully 3D convolutional architectures. While the use of range information helps to filter out a significant portion of motion noise, it cannot entirely suppress interference signals or residual energy from nearby targets. As a result, 3D CNNs which lack explicit mechanisms for temporal modeling face difficulty distinguishing gesture from background noise/interference. In contrast, temporal models (such as CNN-GRU and ViViT) demonstrate increased resilience to real-world imperfections.

\section{Conclusion}
\label{sec:conclusion}
This work introduces WiRD-Gest, a Wi-Fi sensing system that extracts RD maps on a single, unmodified COTS laptop for robust gesture recognition. We provide the first comprehensive benchmark of deep learning models on RD clips, evaluating performance across in-domain, cross-user, and challenging public-space scenarios.Experimental results show that high generalization accuracy is achievable, with CNN-GRU reaching 90\% even in real-world, interference-rich environment. Specifically, we demonstrate how the addition of the range dimension fundamentally shifts the optimal learning approach toward models with explicit temporal tracking. In the future we plan to make publicly available the created dataset, extend the work to automatically segment gestures from streaming Wi-Fi CSI data and further optimize the classification models. 



\bibliographystyle{IEEEbib}
\bibliography{references}

\end{document}